\documentclass[12pt]{article}
\usepackage{graphicx}
\begin{document}
\pagestyle{empty}
%
%%%%%%%%%%%%%%%%%%%%%%%%%%%%%    TITLE   %%%%%%%%%%%%%%%%%%%%%%%%%%%%%%%
%
\begin{center}
{\Large\bf INTERACTING PION PAIRS \\
           IN NUCLEAR MATTER}
\end{center}
%
%%%%%%%%%%%%%%%%%%%   AUTHORS  AND INSTITUTIONS  %%%%%%%%%%%%%%%%%%%%%%
%      
\vspace{-1.0cm}
\begin{center}
\normalsize
N. GRION \\
\small{\it Istituto Nazionale di Fisica Nucleare, 34127 Trieste, Italy} \\ 
\large{The CHAOS Collaboration} 
\end{center}
\vspace{-0.7cm}
%      
%%%%%%%%%%%%%%%%%%%%%%%%%%%%    ABSTRACT   %%%%%%%%%%%%%%%%%%%%%%%%%%%%
%
%
\hspace{6.0cm}{\bf Abstract} \\
\setlength{\baselineskip}{2.3ex}         % double spacing between lines
{\small
The pion-production $\pi^+ A \rightarrow \pi^+\pi^{\pm} A'$ reactions 
were studied on nuclei $^{2}H$, $^{12}C$, $^{40}Ca$ and $^{208}Pb$ at 
a pion energy of $T_{\pi^{+}}$=283 MeV using the CHAOS spectrometer.
The experimental results are reduced to differential cross sections 
and compared to both theoretical predictions and reaction phase space. 
Near the $2m_{\pi}$ threshold pion pairs couple to $(\pi\pi)_{I=J=0}$ 
when produced in the $\pi^+\rightarrow \pi^+\pi^-$ reaction channel.  
The $\cal C$$_{\pi\pi}^A$ ratio between the $\pi^{+}\pi^{\pm}$ invariant 
masses on nuclei and on the nucleon is also presented. The marked 
near-threshold enhancement of $\cal C$$_{\pi^+\pi^-}^A$ is consistent 
with theoretical predictions addressing the partial restoration of 
chiral symmetry. On the opposite, nuclear matter weakly influences 
$\cal C$$_{\pi^+\pi^+}^A$.
}
\normalsize
%
%%%%%%%%%%%%%%%%%%%%%%%%%%%%%%  TEXT  %%%%%%%%%%%%%%%%%%%%%%%%%%%%%%%
%

\vspace{-.2cm}
{\bf 1 Introduction}
\vspace{-.5cm}

The influence of the nuclear medium on the $\pi\pi$ interaction was 
studied at TRIUMF by means of the pion induced pion-production reaction 
$\pi^+A \rightarrow \pi^+\pi^{\pm}A'$ (henceforth labelled $\pi 2\pi$). 
The initial study was directed to the deuterium, to understand the 
$\pi 2\pi$ behaviour on both a neutron and a proton then on complex nuclei 
$^{12}C$, $^{40}Ca$ and $^{208}Pb$ to derive possible $\pi\pi$ 
medium modifications by direct comparison of the $\pi 2\pi$ data. 
\vspace{-.5cm}

Early $\pi 2\pi$ measurements\cite{Camerini:one} found that the near-threshold 
behaviour of the $\pi^+\pi^-$ invariant mass $M_{\pi^+\pi^-}^A$ 
increasingly peak toward the $2m_\pi$ threshold as the nucleus mass 
number increases. Such a behaviour was explained by a theoretical
approach of \cite{Schuck:one} which considered a $\pi\pi$ pair a strongly 
interacting system when coupled to the I=J=0 quantum numbers. The theory 
studies the $(\pi\pi)_{I=J=0}$ properties in nuclear matter by dressing the 
single-pion propagator to account for the $P-$wave coupling of pions 
to $p-h$  and $\Delta-h$ configurations. The model is able to explain 
the general features of the $M_{\pi^+\pi^-}^A$ distributions, which are 
predicted to increasingly accumulate strength near the $2m_{\pi}$ threshold 
for $\rho$, the nuclear medium density, approaching $\rho_0$, the 
saturation density. Within the same theoretical framework, however, 
the near-threshold strength of the $\pi\pi$ $T-$matrix is considerably 
reduced when the $\pi\pi$ interaction is constrained to be chiral 
symmetric\cite{Aouissat:one}, which may indicate that effects other 
than the in-medium $(\pi\pi)_{I=J=0}$ interaction contribute to the 
observed strength. Conversely, the absence of any in-medium modification 
of the $(\pi\pi)_{I=J=0}$ interaction leads to $M_{\pi^+\pi^-}^A$ 
distributions which lack of strength at threshold\cite{Camerini:one,Oset:one}. 
In recent theoretical works on the $(\pi\pi)_{I=J=0}$ interaction in nuclear 
matter\cite{Aouissat:two}, the effects of standard many-body correlations 
are combined with those deriving from the restoration of chiral symmetry in 
nuclear matter. As a result, the $M_{\pi^+\pi^-}^A$ distributions are shown 
to regain strength near the $2m_{\pi}$ threshold as $A$ (thus the average 
$\rho$) increases. Such a property was earlier outlined in some theoretical 
works, which demonstrated that the $M_{\pi^+\pi^-}^A$ enhancement near 
threshold is a distinct consequence of the partial restoration of the chiral 
symmetry at $\rho < \rho_0$ \cite{Hatsuda:one}. The purpose of this 
contribution is to present a set of $\pi 2\pi$ data and to discuss it 
to the light of the most recent theoretical findings.

\vspace{-.3cm}
{\bf 2 The experiment} 
\vspace{-.5cm}

The experiment was carried out at TRIUMF. Incident pions were produced 
by the collision of 480 MeV protons on a 10 mm thick graphite target. The 
M11 pion beam line transported the 282.7 MeV pions to the final focus. 
Pion pairs were detected in coincidence to ensure an unique identification 
of the pion production process. 
\vspace{-.5cm}

CHAOS is a magnetic spectrometer which was designed for the detection 
of multi-particle events in the medium-energy range \cite{CHAOS:one}. 
The magnetic field is generated by a dipole whose pole tip has 95 cm 
in diameter. The magnet is capable of producing a field intensity up 
to 1.6 T with an uniformity of about 1\%. The top side has a 12 cm 
bore at the centre for the insertion of targets. Fig. 1 illustrates 
\begin{figure}[h]
 \centering
  \includegraphics*[angle=90,width=0.6\textwidth]{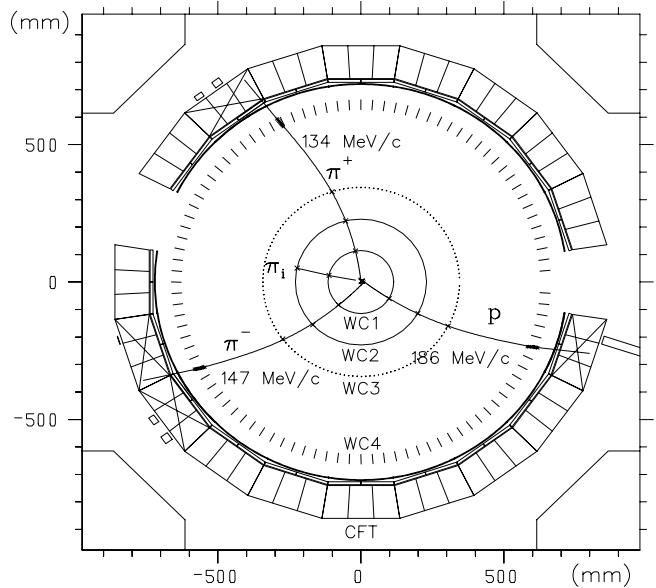}
  \setlength{\abovecaptionskip}{7pt}  % default = 10 pt
  \setlength{\belowcaptionskip}{0pt}  % default = 0pt!
  \caption{\footnotesize Reconstructed particle trajectories in CHAOS
    for the $\pi^+_i \rightarrow \pi^+\pi^-p$ reaction on $^{12}C$, 
    the geometrical disposition of the wire chambers (WC), the first 
    level trigger hardware (CFT) and the magnet return yokes in the 
    corners. Two CFT segments are removed to permit the particle beam 
    ($\pi_i$) to traverse the spectrometer. The CFT segments which are 
    hit by particles are marked with crosses, and the energy deposited
    in $\Delta E1$ and $\Delta E2$ is indicated by boxes. The proton 
    has a momentum  slightly above the CHAOS threshold (185 MeV/c),
    and its energy is fully deposited in $\Delta E1$.}
\end{figure}
a reconstructed $\pi^+_i \rightarrow \pi^+\pi^-p$ events on $^{12}C$ 
and the geometrical disposition of the wire chambers (WC), the CHAOS 
first level trigger hardware (CFT) and the magnet return yokes in the 
corners. WC1 and WC2 are multiwire proportional chambers which are 
capable of handling rates exceeding 5$\times 10^6$ particles/s for 
extended periods of time with high efficiency ($\sim$95\%). They have 
a cylindrical shape with a diameter of 22.8 cm and 45.8 cm, respectively. 
WC3 is a cylindrical drift chamber designed to operate in a magnetic 
field, the chamber diameter is 68.6 cm. The outermost 
chamber WC4 is a vector drift chamber 122.6 cm in diameter, which 
operates outside of the magnetic field of CHAOS. The segments of WC3 
and WC4 which were crossed by the incident particle beam were turned 
off. The CFT hardware consists of three coaxial cylindrical layers of 
fast-counting detectors. The first two layers are 
NE110 plastic scintillators 0.3 cm and 1.2 cm thick, respectively. 
$\Delta E1$ is 72 cm far from the magnet centre and spans a zenith angle 
of $\pm 7^{\circ}$; thus, it defines the geometrical solid angle of CHAOS 
$\Omega$=1.5 sr. The third layer is a SF5 lead-glass 12.5 cm thick, 
about 5 radiation lengths, which is used as a Cerenkov counter. The 
three layers were segmented in order to provide an efficient triggering 
system to multi-particle events. Each segment covered an azimuthal 
angle (i.e., in-the-reaction plane) of 18$^\circ$. 

\vspace{-.3cm}
{\bf 3 Analysis} 
\vspace{-.5cm}

The data reduction was based on fully reconstructed 
$\pi^+ \rightarrow \pi^+\pi^-$ and $\pi^+ \rightarrow \pi^+\pi^+$  
events. In order to form differential cross sections these events 
were binned with their weights, which include the $\pi^+\pi^{\pm}$
decay rates inside CHAOS. The capability of the CHAOS spectrometer 
of measuring the 
kinetic energies ($T$) and laboratory angles ($\theta$) of 
each $\pi \rightarrow \pi_1\pi_2$ event permits the 
determination of the five-fold differential cross section 
$\partial^5\sigma/(\partial T\partial\Theta)_{\pi_1}
(\partial T\partial\Theta)_{\pi_2}\partial\Phi_{\pi_1\pi_2}$, 
where $\Phi_{\pi_1\pi_2}$ is the zenithal angle between 
$\pi_1$ and $\pi_2$, which CHAOS enabled measurement at both 
$180^{\circ}\pm 7^{\circ}$ and $0^{\circ}\pm 7^{\circ}$.
Four-fold differential cross sections were then obtained by 
integrating out the $\Phi_{\pi\pi}$ dependence, which was 
performed by using a linear function joining the two measured
data points. Such an assessment of the $\Phi_{\pi\pi}$ 
dependence reflected in a systematic uncertainty of 8\% 
($\sigma$) for the deuterium and 6-7\% ($\sigma$) for nuclei. 
Furthermore, the data were reduced to single differential cross 
sections $d\sigma/d{\cal O}_{\pi}$, where ${\cal O}_{\pi}$ 
represents $(T_{\pi}$ or $\Theta_{\pi})_{1,2}$ or a combination 
of them. The cross section was then related 
to measured quantities $\frac{d\sigma}{d{\cal O}_{\pi}}$=
$f_e\frac{N({\cal O}_\pi)}{\Delta{\cal O}_\pi}$, where $f_e$ is 
a parameter which is determined by the experimental conditions.

\vspace{-.3cm}
{\bf 4 Results of the $\pi \rightarrow \pi\pi$ reaction in nuclei} 
\vspace{-.5cm}

A general property of the $\pi 2\pi$ process on nuclei in the 
low-energy $M_{\pi\pi}$ regime was outlined by previous experimental 
works: it is a quasi-free process both when it occurs on deuterium
\cite{Rui:one} and on complex nuclei\cite{Bonutti:two}. Furthermore, 
a common reaction mechanism underlies the process whether it occurs 
on a nucleon or a nucleus\cite{Bonutti:four}. Thus the study of the 
$\pi^+$$^2H\rightarrow\pi^+\pi^{\pm}NN$ reaction is dynamically 
equivalent to studying the elementrary $\pi^+ n \rightarrow \pi^+\pi^- p$ 
and $\pi^+ p \rightarrow \pi^+\pi^+ n$ reactions separately.
\begin{figure}[t]
 \centering
  \includegraphics*[angle=0,width=0.50\textwidth]{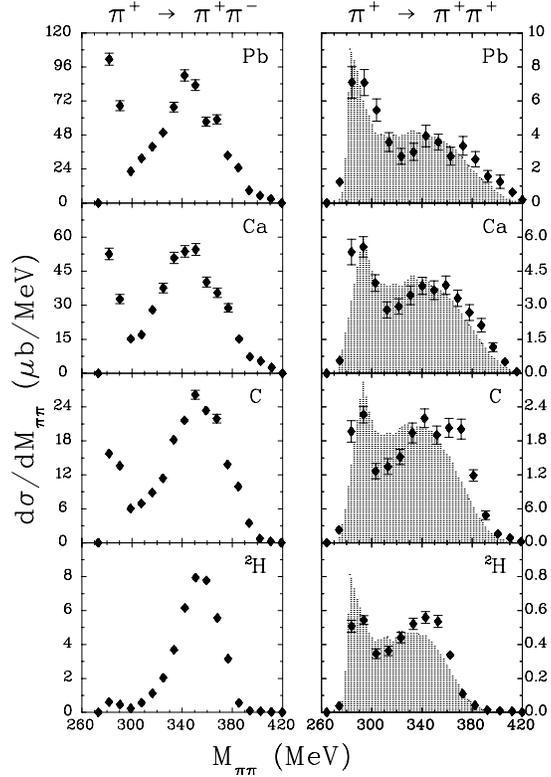}
  \setlength{\abovecaptionskip}{7pt}  % default = 10 pt
  \setlength{\belowcaptionskip}{0pt}  % default = 0pt!
  \caption{\footnotesize Invariant mass distributions (diamonds) for 
      the $\pi^+ \rightarrow \pi^+\pi^-$ and $\pi^+ \rightarrow \pi^+\pi^+$ 
      reactions on $^{2}H$, $^{12}C$, $^{40}Ca$ and $^{208}Pb$.
      Diagrams (dots) are the result of phase-space simulations for the
      pion-production $\pi A \rightarrow \pi\pi N [A-1]$ reaction.} 
\end{figure}
In the present measurement, the $\pi^+ A\rightarrow \pi^+\pi^{\pm} A'$ 
reactions were studied under the same experimental conditions. Thus
for a given observable the distributions are directly comparable. 
In addition, the moderate out-of-the-reaction plane angular acceptance 
of CHAOS may condition the intrinsic shape of the distributions. They 
are not corrected for it. The error bars explicitly reported on the 
spectra are the statistical uncertainties. 
\begin{figure}[t]
 \centering
  \includegraphics*[angle=0,width=0.52\textwidth]{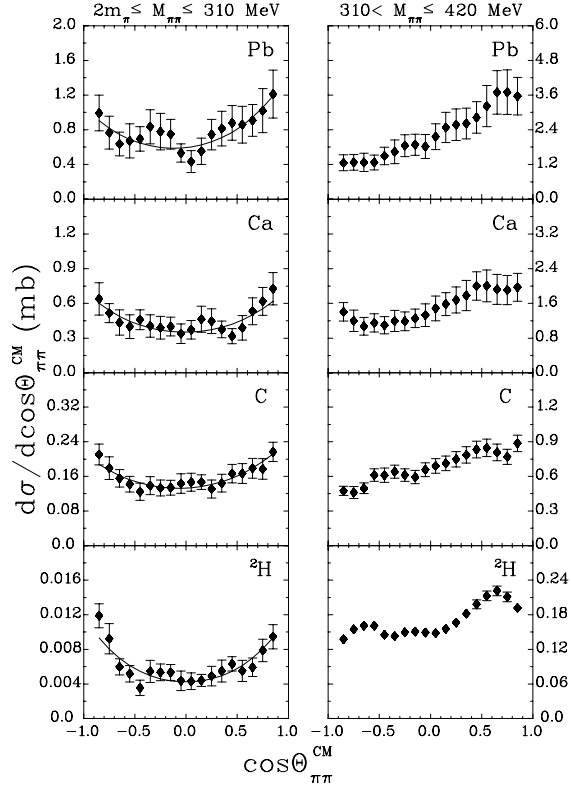}
  \setlength{\abovecaptionskip}{7pt}  % default = 10 pt
  \setlength{\belowcaptionskip}{0pt}  % default = 0pt!
  \caption{\footnotesize Distribution of the cos$\Theta$ 
    in the $\pi^+\pi^-$ centre-of-mass frame for the 
    $\pi^+ A \rightarrow \pi^+\pi^- A'$ reaction at
    $2m_{\pi}\leq M_{\pi\pi}\leq 310 MeV$ (left frame), and 
    at $310 < M_{\pi\pi}\leq 420 MeV$ (right frame). The
    solid lines are best-fit to the data including S, 
    P and D waves. }
\end{figure}
\vspace{-.5cm}

Fig. 2 shows the single differential cross sections (diamonds) as 
a function of the $\pi\pi$ invariant mass ($M_{\pi\pi}$, MeV) for the 
two $\pi^+ \rightarrow \pi^+\pi^-$ and $\pi^+ \rightarrow \pi^+\pi^+$ 
reaction channels. The horizontal error bars are not indicated since 
they lie within symbols. The distributions span the energy interval 
available to the $\pi 2\pi$ reaction which ranges from $2m_{\pi}$, 
the low-energy threshold, up the 420 MeV, the maximum allowed by the 
reaction. The $\pi A \rightarrow \pi\pi N [A-1]$ phase space simulations 
(dotted histograms) are also provided and are normalized to the area 
subtended by the experimental distributions. 
\vspace{-.5cm}

Regardless of the nucleus mass number, the invariant mass for the 
$\pi^+ \rightarrow \pi^+\pi^+$ distributions closely follow phase 
space and the energy maximum increases with the increase of $A$, 
that is, with the increase of the nuclear Fermi momentum. The 
$\pi^+\rightarrow\pi^+\pi^-$ channel discloses a different behaviour; 
as compared to phase space, the $^{2}H$ invariant mass displays little 
strength from $2m_{\pi}$ to 310 MeV while, on the same energy interval, 
the $^{12}C$, $^{40}Ca$ and $^{208}Pb$ $\pi^+\pi^-$ invariant mass 
distributions increasingly peak as $A$ increases. 
\vspace{-.5cm}

In order to explain the nature of the reaction mechanism contributing 
to the peak structure, it is useful to examine the cos$\Theta_{\pi\pi}^{CM}$ 
distribution in the invariant mass interval of the peak, where 
$\Theta_{\pi\pi}^{CM}$ is the angle between the direction of a final pion 
and the direction of the incoming pion beam in the $\pi^+\pi^-$ rest frame. 
Fig. 3 shows the cos$\Theta_{\pi\pi}^{CM}$ distributions (diamonds) for 
$2m_{\pi}\leq M_{\pi^+\pi^-}\leq 310$ MeV and $310 < M_{\pi^+\pi^-}\leq 420$ 
MeV, the latter being shown for comparison. The vertical error bars are 
the overall uncertainties, which sum in quadrature the systematic and the 
statistic uncertainties. The differential cross sections are best-fitted 
(solid line) with a partial wave expansion limited to the three lowest waves, 
i.e. S, P and D. For all the studied nuclei is $\chi^2_{\nu}\leq 1$ which 
indicates that a proper number of waves was used in the expansion. In the 
case of heavier nuclei, the $\pi^+\pi^-$ system predominantly couples 
$S-$wave $\sim$ 95\% and a remaining 5\% is spent in a $D-$wave state. 
Furthermore, within the sensitivity of the $\chi^2_{\nu}-$method any 
$P-$wave coupling of the two pions is excluded. 
\vspace{-.5cm}

Two recent theoretical works \cite{Rapp:one,Vicente:one} have modelled 
the $\pi 2 \pi$ reaction on nuclei whose results are reported in Fig. 4. The 
(short and long) dashed lines denote the calculations of\cite{Rapp:one}
while the results of the predictions are shown with the full lines
\cite{Vicente:one}. In the case of $Ca$, $R1$ ($R2$) indicates the predictions 
for $\rho$=0.7$\rho_n$ ($\rho$=0.5$\rho_n$), where $\rho_n$ is the nuclear 
saturation density, while $V1$ is the result of the calculations for a mean 
$\rho$=0.24$\rho_n$\cite{Vicente:two}. For both $^{2}H$ and $^{40}Ca$ the 
curves are normalized to the experimental data. For the 
$\pi^+ \rightarrow \pi^+ \pi^+$ channel, the predictions agree with the 
invariant mass distributions. The $R1$ and $R2$ distributions slightly differ 
from each other: $R1$ presents a broader shape which is due to the larger 
nuclear Fermi momentum as a consequence of the higher nuclear density 
used\cite{Rapp:one}. The lower $\rho$ used for $R1$ seems to better fit the 
measured distribution, which is a trend also supported by $V1$. Therefore, 
both models and indicate that the average nuclear density of $^{40}Ca$ for 
the production reaction to take place cannot exceed 0.5$\rho_n$.
\vspace{-.5cm}
\begin{figure}[t]
 \centering
  \includegraphics*[angle=0,width=0.6\textwidth]{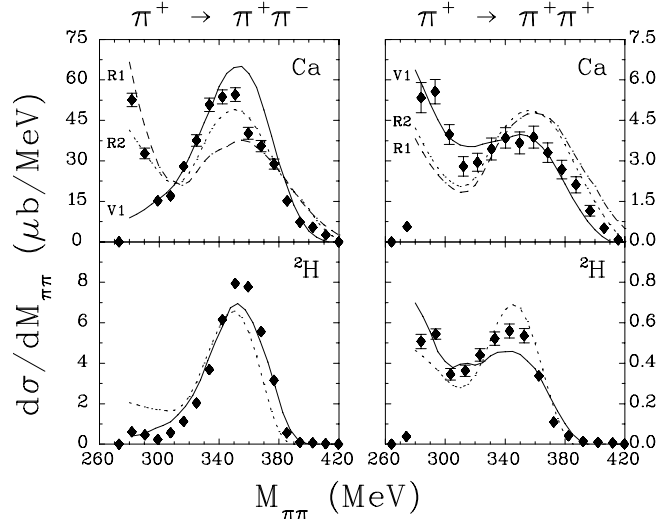}
  \setlength{\abovecaptionskip}{7pt}  % default = 10 pt
  \setlength{\belowcaptionskip}{0pt}  % default = 0pt!
  \caption{\footnotesize Invariant mass distributions (diamonds) for 
      the $\pi^+ \rightarrow \pi^+\pi^-$ and $\pi^+ \rightarrow \pi^+\pi^+$ 
      reactions on $^{2}H$ and $^{40}Ca$. The dashed curves are taken from 
      Ref.\cite{Rapp:one}, $R1$ and $R2$ are the results for 0.7$\rho_n$
      and 0.5$\rho_n$, respectively. The full curves ($V1$) are from the
      theoretical work of \cite{Vicente:one} for a mean density
      $\rho$=0.24$\rho_n$.}
\end{figure}

In the $\pi^+ \rightarrow \pi^+ \pi^-$ channel the distribution predicted 
by\cite{Vicente:one} for $^{2}H$ is able to describe the data, 
although it tends to overestimate the low-energy $M_{\pi\pi}$ yield. 
The present version of the model surely improves the previous 
versions\cite{Oset:one}, thus allowing for the construction
of nuclear medium effects on a reliable ground. The approach of 
\cite{Vicente:one} includes several medium effects: Fermi motion, 
pion absorption, pion quasi-elastic scattering and $(\pi\pi)_{I=J=0}$
medium modifications, which are able to reproduce only a moderate 
$M_{\pi\pi}$ strength in the near-threshold region. In addition, an
increase of $\rho$ from 0.24$\rho_n$ to 0.5$\rho_n$ and to 0.7$\rho_n$ 
is unlikely to improve the agreement with the experimental cross section,
see also Fig. 8 of Ref.\cite{Vicente:one}. In the case of\cite{Rapp:one}, 
the $R2$ prediction (0.5$\rho_n$) seems to better reproduce the $M_{\pi\pi}$ 
distribution, although some of the near threshold yield already derives 
from the the low-energy $\pi^+$ $^2H \rightarrow \pi^+ \pi^- pp$ production. 
Therefore, $V1$, $R1$ and $R2$ suggest that the $M_{\pi^+\pi^-}$ missing 
strength near the 2m$_\pi$ threshold should be searched in a stronger 
$\rho$-dependence of the $(\pi\pi)_{I=J=0}$ interaction, rather than 
requiring an unlikely high-density nuclear environment for the 
pion-production process to occur.
\vspace{-.5cm}
\begin{figure}[t]
 \centering
  \includegraphics*[angle=90,width=0.7\textwidth]{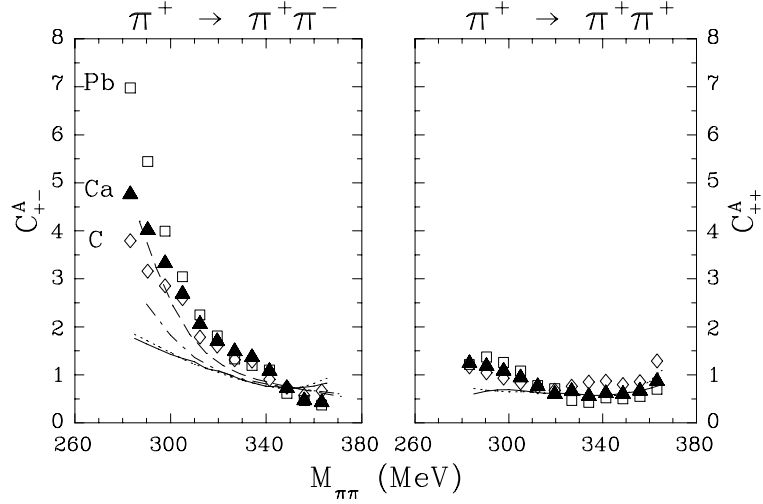}
  \setlength{\abovecaptionskip}{7pt}  % default = 10 pt
  \setlength{\belowcaptionskip}{0pt}  % default = 0pt!
  \caption{\footnotesize The composite ratios $\cal C$$_{\pi\pi}^A$
      for $^{12}C$, $^{40}Ca$ and $^{208}Pb$. The curves are taken from 
      \cite{Vicente:one} (full), \cite{Rapp:one} (dotted), \cite{Hatsuda:one} 
      (dash-dotted) and \cite{Aouissat:two} (dashed). Further details are 
      reported in the text.}
\end{figure}

The observable $\cal C$$_{\pi\pi}^A$ is presented 
in comparison with recent theoretical predictions. $\cal C$$_{\pi\pi}^A$ 
is defined as the composite ratio $\frac{M_{\pi\pi}^A}{\sigma_T^A} / 
\frac{M_{\pi\pi}^N}{\sigma_T^N}$, where $\sigma_T^A$ ($\sigma_T^N$) 
is the measured total cross section of the $\pi 2\pi$ process in nuclei 
(nucleon). This observable has the property of yielding the net effect 
of nuclear matter on the $(\pi\pi)_{I=J=0}$ interacting system regardless 
of the $\pi 2\pi$ reaction mechanism used to produce the pion pair
\cite{Bonutti:four}. Therefore, $\cal C$$_{\pi\pi}^A$ can be compared 
with the \cite{Rapp:one,Vicente:one} predictions which explicitly 
calculate both $M_{\pi\pi}^{Ca}$ and $M_{\pi\pi}^{^{2}H}$, but also with 
the theories described in \cite{Aouissat:two,Hatsuda:one} because they 
calculate the mass distribution of an interacting $(\pi\pi)_{I=J=0}$ 
system both in vacuum and in nuclear matter. Since 
the above calculations are reported either in arbitrary units
\cite{Rapp:one,Vicente:one} or in units which are complex to scale
\cite{Aouissat:two,Hatsuda:one}, theoretical predictions are normalized 
to the experimental distributions at $M_{\pi\pi}$=350$\pm$10 MeV, where 
$\cal C$$_{\pi\pi}^A$  presents a flat behaviour Fig. 5.
\vspace{-.5cm}

For both reaction channels, the full\cite{Vicente:one} and dotted
\cite{Rapp:one} curves in Fig. 5 are obtained by simply dividing 
$M_{\pi\pi}^{Ca}$/ $M_{\pi\pi}^{^{2}H}$. Furthermore, for both 
approaches the underlying medium effect is the $P-$wave coupling of 
$\pi$'s to $p-h$ and $\Delta-h$ configurations, which accounts for the 
near-threshold enhancement. When applied to the $\cal C$$_{\pi\pi}^{Ca}$, 
both \cite{Rapp:one} and \cite{Vicente:one} predict the same result; 
in fact, they well describe the behaviour of $\cal C$$_{++}^{Ca}$ 
throughout the $M_{\pi\pi}$ energy range, while for $\cal C$$_{+-}^{Ca}$ 
only part of the near-threshold strength is reproduced. The models of 
Refs. \cite{Aouissat:two,Hatsuda:one} examine the medium 
modifications on the scalar-isoscalar meson, the $\sigma-$meson. Nuclear 
matter is assumed to partially restore chiral symmetry and consequently 
$m_{\sigma}$ to vary with $\rho$, the variation being parametrised as 
$1-p\frac{\rho}{\rho_n}$, where $p$ can range in the interval 
0.1$\leq p \leq$0.3 for\cite{Hatsuda:one}  and 0.2$\leq p\leq$0.3 for 
\cite{Aouissat:two}, and for both is $\rho$=$\rho_n$. Both models are 
capable of yielding large strength 
near the $2m_\pi$ threshold, therefore the capability of predicting 
$\cal C$$_{+-}^A$ is compared for a common minimum value of the 
parameter $p$=0.2. In Fig. 5 the predictions of \cite{Hatsuda:one}  and 
\cite{Aouissat:two} are reported with dash-dotted line and dashed line, 
respectively. The model \cite{Aouissat:two} of provides a larger 
near-threshold strength, which is due to the combined contributions of 
the in-medium $P-$wave coupling of pions to $p-h$ and $\Delta-h$ 
configurations and to the partial restoration of chiral symmetry in 
nuclear matter. This model, however, is still too schematic for a 
conclusive comparison to the present data, therefore full theoretical 
calculations are called for.

\vspace{-.3cm}
{\bf 5 Conclusions} 
\vspace{-.5cm}

$\cal C$$_{\pi\pi}^A$ was found to yield the net effect of nuclear matter 
on the $\pi\pi$ system regardless of the $\pi 2\pi$ reaction mechanism 
used to produce the pion pair. These distributions display a marked 
dependence on the charge state of the final pions: 
  (i) the $\cal C$$_{\pi^+\pi^-}^A$ distributions peak at the $2m_{\pi}$ 
      threshold and the yield increases as $A$ increases thus denoting 
      that pion pairs form a strongly interacting system; furthermore, the 
      $\pi\pi$ system couples to the $I=J=0$ channel, the $\sigma -$meson 
      channel. 
 (ii) In the $\pi^+ \rightarrow \pi^+\pi^+$ channel, the $\cal C$$_{\pi\pi}^A$ 
      behaviour barely depends on both $A$ and $T$ thus indicating that 
      nuclear matter weakly affect the $(\pi\pi)_{I,J=2,0}$ interaction. 
The $\cal C$$_{\pi^+\pi^-}^A$ observable was compared with theories studying 
the $(\pi\pi)_{I=J=0}$ in-medium modifications associated to the partial 
restoration of chiral symmetry in nuclear matter, and with model calculations
which only include standard many-body correlations, i.e. the $P-$wave 
coupling of $\pi's$ to $p-h$ and $\Delta-h$ configurations. It was found 
that both mechanisms are necessary to interpret the data, although chiral 
symmetry restoration yields the larger near-threshold contribution. Whether 
this conclusion is correct, the $\pi 2\pi$ CHAOS data would indicate an
example of a distinct $QCD$ effect in low-energy nuclear physics. 
\vspace{-.5cm}

Montecarlo simulations of the $\pi^+ A \rightarrow \pi^+\pi^{\pm} N [A-1]$ 
reaction phase space revealed useful to interpret some of the $\pi 2\pi$ 
data. In the case of $M_{\pi^+\pi^+}$, $\pi^+\pi^+$ pairs distribute 
according to phase space. In addition, simulations are able to describe 
the high-energy part of the distributions which are sensitive to the nuclear 
Fermi momentum of the interacting $\pi^+ p[A-1] \rightarrow \pi^+\pi^+ 
n[A-1]'$ proton. For the $M_{\pi^+\pi^-}$ distributions the $\pi\pi$ dynamics 
overwhelms the dipion kinematics: unlike phase space, the near-threshold 
$\pi^+\pi^-$ yield is suppressed in the elementary production reaction, 
$\pi^+$ $^2H \rightarrow \pi^+\pi^- pp$ in the present work, while in the 
same energy range medium modifications strongly enhance $M_{\pi^+\pi^-}$. 
A guideline to the interpretation of the $M_{\pi^+\pi^{\pm}}^A$ behaviour 
should combine the effects of the chiral symmetry restoration in nuclear 
matter and standard many-body correlations. Such an approach would exclude 
high-density nuclear matter for both the production reaction to take place 
and the $\pi\pi$ system to undergo medium modification.
\vspace{-1.0cm}
%
%%%%%%%%%%%%%%%%%%%%%%%%%%%  THE BIBLIOGRAPHY %%%%%%%%%%%%%%%%%%%%%%%%%%
%
 
\end{document}